# Secondary Analysis of Teaching Methods in Introductory Physics: a 50k-Student Study

Joshua Von Korff*

*Department of Physics and Astronomy, Georgia State University, Atlanta, GA 30303*

Benjamin Archibeque and K. Alison Gomez

*Department of Physics, Kansas State University, Manhattan, KS 66506*

Tyrel Heckendorf

*Department of Physics and Astronomy, Georgia State University, Atlanta, GA 30303*

Sarah B. McKagan

*American Association of Physics Teachers, College Park, MD 20740*

Eleanor C. Sayre and Edward W. Schenk

*Department of Physics, Kansas State University, Manhattan, KS 66506*

Chase Shepherd

*Department of Physics and Astronomy, Georgia State University, Atlanta, GA 30303*

Lane Sorell

*Department of Physics, Kansas State University, Manhattan, KS 66506*

## Abstract

Physics education researchers have developed many evidence-based instructional strategies to enhance conceptual learning of students in introductory physics courses. These strategies have historically been tested using assessments such as the Force Concept Inventory (FCI) and Force and Motion Conceptual Evaluation (FMCE). We have performed a review and analysis of FCI and FMCE data published between 1995 and 2014. We confirm previous findings that interactive engagement teaching techniques are significantly more likely to produce high student learning gains than traditional lecture-based instruction. We also establish that interactive engagement instruction works in many settings, including those with students having a high and low level of prior knowledge, at liberal arts and research universities, and enrolled in both small and large classes.

## I. INTRODUCTION

Over the last three decades, widespread reform efforts in physics courses have transformed physics instruction throughout the world.[1-4] These efforts have emerged from the growing field of Physics Education Research (PER), and have been highly dependent on carefully designed research-based assessments of conceptual understanding, including both open-ended assessments[5] and multiple-choice assessments.[6] Without well-designed assessments, instructors and researchers cannot ascertain whether their efforts are effective, and without shared ones, they cannot compare their efforts. Conceptual assessments are especially important, since studies have shown that many students have trouble with conceptual questions even when they perform well on quantitative tests of problem-solving.[7]

Researchers have repeatedly found that diverse "interactive engagement" (IE) instructional techniques have a great advantage over traditional lecture-based instruction (TRAD).[8,9] Students can learn more physics when they actively participate in physics discussion and experimentation in the classroom. To test the efficacy of these interactive engagement techniques, nearly 50 multiple-choice conceptual assessments have been developed across dozens of physics topics.[6,10] While open-ended assessments can provide much deeper insight into the details of student thinking and are often used to refine and assess a particular teaching method, easily-graded multiple-choice assessments are more useful for providing a standard that can be administered to large numbers of students across a wide variety of settings to assess the impact of interactive engagement teaching in general. For example, Hake[9] found dramatic differences between IE and TRAD in analysis of scores for 6000 students on the Force Concept Inventory (FCI),[11] across a wide variety of institutions. The FCI is the most highly-cited conceptual assessment across all of discipline-based education research, with 2479 citations. The FCI tests students' understanding of Newtonian forces and motion, and it is appropriate for introductory physics students at the university level. The Force and Motion Conceptual Evaluation (FMCE)[12,13] is another popular and related conceptual assessment with 610 citations. The FCI and FMCE are strongly correlated, but may give differing evaluations of a particular group of students or instructional technique.[14]

In the 18 years since the publication of the Hake study, the quantity of published data on research-based conceptual assessments of mechanics has increased by an order of magnitude. The current study confirms the results of earlier studies with a much larger sample.

## II. METHODS

We conducted a secondary analysis on published data about FCI and FMCE scores and gains. We conducted a literature search for peer-reviewed papers and conference proceedings from 1992 (when the FCI was published) to 2014 using ERIC, Scopus, and Web of Science using the keywords "FCI", "FMCE", "Force Concept Inventory", and "Force and Motion Conceptual Evaluation". Additionally, we examined every PER-related paper in three major physics education journals: *Physical Review Special Topics – Physics Education Research*; the *American Journal of Physics*; and the *Physics Education Research Conference Proceedings*. From the papers, we collected information about teaching methods, class sizes, and student performance. We enriched this data with information from the Carnegie Classification database,[15] which categorizes different kinds of institutions, and with data on average SAT scores at each institution from the College Board's College Handbook using editions from 1994, 1999, 2004, 2009, and 2014.[16] All of the data used for the secondary analysis are available online.[17] By collecting all published data from the two most widely used assessments in physics education, we intend to shed light on the factors that have improved students' learning gains.

To be included, studies needed to conform to all of the following three criteria. First, students must have been enrolled in introductory college or university mechanics courses (not high school students or teachers) in the United States or Canada. Second, the authors must report students' (aggregate) scores from the full FCI or FMCE using a standard scoring scheme (not a subset or variant of the test). Lastly, the data must compare students at the beginning and end of the course, by reporting average pre and post scores and/or normalized gain (described below). Where multiple papers reported the same data, we removed the duplicates. Where data was reported only broken out by gender or other factors, we have combined them using a weighted average. Seventy-two papers survived our selection criteria, representing about 600 classes and about 45,000 students. Breaking this down further, 63(15) papers reported data from the FCI(FMCE), representing about 450(150) classes and about 31,000(14,000) students. At least two researchers examined each data element in our database.

There are several metrics for comparing students' learning at the beginning and end of a course. By far, the most popular in physics education research is the "normalized gain" (also referred to as "gain").[9] This is the difference between the scores at the beginning and at the end (the "raw gain") divided by the number of problems they got wrong at the beginning (the "normalizing factor"). It is possible to compute a classroom gain either using the average pre-test and post-test scores, or by finding each student's gain and then taking the average. Hake initially defined normalized gain using the former approach,[9] but many researchers use the latter approach, and Hake points out that the two approaches do not normally differ by more than 5%. We permitted both approaches, since both are widely used and many authors did not state which approach they used.

In social sciences research outside of physics, it is more common to report an effect size[18] than a gain. Recent meta-analyses of student learning across STEM disciplines have compared various teaching methods and scores on different instruments using effect size.[19] These studies however, must discard most of the available data because the original authors reported gain, not effect size. While we agree that effect size is a stronger statistic, we take a practical approach: the consensus in physics education research over the last 20 years has been to report gain only, so gain is how we compared studies.

Note that, like Hake, we chose the individual class as the unit of analysis for our calculations, since this is the smallest unit of analysis that is commonly available in the published literature, and the teaching method is generally the same for an entire class, but not for different classes in the same department or even by the same instructor. Thus, when we report an average, we mean that we have averaged over all classes in our database, including sometimes over multiple classes from the same university. When an author reported an average gain (or pre-test or post-test score) for multiple classes, but no individual gains, we counted each class as having the average gain.

## III. RESULTS

Our analysis confirms previous studies[8,9] showing that classes that use interactive engagement have significantly higher learning gains than those that use traditional lecture (Fig. 1). There is large variation in gains within both traditional lecture and interactive engagement (Fig. 2), suggesting that there are other factors that impact gains. Class size does not impact gains (Fig. 3), nor does the institution type (Fig. 4) or the incoming SAT score of students at the institution, (operationalized by both the institution's 25th and 75th percentile scores for both the math and verbal SAT components). In support of some previous studies[9] and in contrast to others,[20] we find that gains do not correlate with pre-test scores after correcting for teaching method and test. This suggests that gain is a good measure for differentiating between teaching methods and is independent of students' initial preparation, at least when looking at class averages.

**A. How does gain vary by teaching method?**

Figure 1 shows that gains are significantly higher for interactive engagement instruction than for traditional lecture-based instruction on both the FCI and the FMCE. This result was expected, but it is worth emphasizing, since TRAD is still very popular in spite of the weight of evidence against it. In our data set, the FMCE was more likely to discriminate between IE and TRAD, in the sense of assigning higher gains to IE and lower scores to traditional instruction. Gains on the two tests are not directly comparable because they are measuring slightly different things.[14]

**B. What is the variance in gain for a given teaching method?**

Figure 2 shows that there is a substantial variation in normalized gain within both the traditional and interactive categories. This variation suggests that teaching method is not the only factor leading to student learning. We suspect that the variation can be explained by variation in the quality of implementation of IE methods and/or the amount time spent on these methods. We were unable to reliably quantify either of these factors in our analysis of the published literature. The particular IE methods selected by the instructor may also be a relevant factor. As we show below, this variation *cannot* be explained by class size, type of institution, average institutional SAT score, or average pre-test score.

**C. How does gain vary by class size?**

It is often suggested that small classes lead to more effective instruction.[21] However, our data lend little support to this hypothesis in the case of undergraduate physics courses. For this analysis, we discarded very small (<10 students) or very large (>400 students) classes as probable outliers or artifacts from secondary analysis (<5% of classes). We ran ANOVAs on test, teaching method, and class size to discover if there was a primary effect of class size after test and teaching method. We found none ($p>0.2$).

We believe that instructors of large lecture classes selected instructional strategies that compensated for their class size. The most prominent large-course strategy that emerges from our data is Peer Instruction, a technique which combines a lecture format with individual work and group discussion of conceptual questions, where students vote on the response to the questions and the instructor can immediately gauge and respond to class understanding.[2,22] Peer Instruction was used in 6% of IE classes with less than 30 students, 22% of IE classes with 30-100 students, and 41% of IE classes of more than 100 students. Peer Instruction is scalable to arbitrarily large classrooms and is compatible with a lecture format. Large classes also often use lecture or Peer Instruction in the "lecture" section, and other interactive engagement techniques, such as Tutorials in Introductory Physics[4,23] in the "recitation" section, which is effectively a small class taught by teaching assistants. Small classes were more likely to use other methods that are not suitable for large courses, such as Modeling Instruction [24,25] or Studio Physics.[26]

**D. How does gain vary by institution type?**

Many faculty believe that the kind of institution that students are taught at affects their learning gains. For example, faculty at less selective institutions often think their students cannot possibly have the same gain as those for example at Harvard, because their students have less incoming knowledge.

Conversely, many faculty at small liberal arts schools believe that their teaching surpasses the teaching at large impersonal research universities. We investigated whether a school's Carnegie classification (whether it is an Associates-, Bachelors-, Masters-, or Doctoral-granting institution) [15] affected students' learning gains. We found that pre-test scores correlated with Carnegie classification, but gain did not. We used a multifactor ANOVA to look for effects of test (FCI vs. FMCE), teaching method (IE vs. traditional), and institutional type (Carnegie Classification) on normalized gain, then a Tukey HSD test to look for differences. After controlling for teaching method and test, institutional type does not affect normalized gain. Figure 4 illustrates this result for the FCI. The trend is similar for the FMCE, but there are too many institution types for which there is little or no FMCE data to be able to do meaningful analysis of institution type for the FMCE. Even for the FCI, these data need to be read with some caution: there are few published studies at small liberal arts schools and community colleges, and many at large research universities, so there may be a selection bias in our data.

**E. How does gain vary by average SAT scores?**

A study by Coletta and Phillips[27] found a correlation between students' SAT scores and normalized gain. We looked for a correlation between each university's 25$^{th}$ and 75$^{th}$ percentile math and verbal SAT scores for incoming students[16] and normalized gain. As with Carnegie classification, we found that pre-test scores correlated with SAT scores, but gains did not. We corrected for any possible changes in a university's SAT scores by examining the SAT scores from within two years of the time the students' FCI or FMCE data were collected. We also corrected for overall inflation or deflation of SAT scores.[28] One confounding factor in this kind of analysis is that the university's overall SAT scores are not the same as the SAT scores of the students who take any given introductory physics course at that university.

**F. How does gain vary by average pre-test score?**

Hake initially proposed normalized gain as a useful measure because he found that it did not correlate with pre-test score, and therefore can be used to compare teaching methods across different types of institutions and classes where students have varying levels of preparation. This claim has been questioned by Coletta and Phillips,[20] who observed a correlation *r=0.63* between pre-test score and gain in their study of 38 classes at seven institutions. However, we have a much broader array of participating institutions – about ten times as many; and more than three times as many as Hake and Coletta and Phillips combined. We found no correlation *(p=0.47)* between class-average pre-test scores and normalized gain after correcting for test and teaching method. Coletta and Phillips also stated that within a particular university, individual students' pre-test FCI scores are sometimes correlated with normalized gains, with the correlation coefficient ranging from *r=0.15* to *r=0.33* for three different institutions. (At a fourth institution, there was no significant correlation.) Our analysis, like that of Hake, looks only at class averages, not at individual students' scores or gains like this second analysis of Coletta and Phillips. Thus, while it is possible that correlations exist at the student-level for within-class comparisons, our analysis suggests that pre-test scores do not influence class-average gains across institutions, so gain is a useful measure for comparing classes across institutions.

## IV. CONCLUSION

Overall, our finding is that interactive engagement (IE) instruction results in greater learning gains on the FCI and FMCE than traditional lecture instruction, and that variables including class size, SAT scores, pre-test score, and Carnegie classification are not correlated with gain. This confirms that the normalized gain is a powerful tool for measuring the benefits of IE instruction, and that it can be used to make comparisons between courses taught in a variety of contexts.

The principal limitation of this study is that researchers may report their gains selectively. For instance, they may choose not to publish gains that are below a certain value. However, we found a reported gain that was less than zero, indicating that the students appeared to "unlearn" the concept of force. This suggests that at least some researchers were willing to report low gains. Additionally, if selective reporting equally affects classes of different sizes, SAT scores, pre-test scores, and Carnegie classification, then it would not substantially alter the correlation of gain with these variables.

**ACKNOWLEDGMENTS**

We were funded by the NSF WIDER program (DUE-1256354, DUE-1256354, DUE-1347821, DUE-1347728) and through a Research Experience for Undergraduates (PHYS-1461251).

*Correspondence to: jvonkorff@gsu.edu

1      R. J. Beichner, J. M. Saul, D. S. Abbott, J. J. Morse, D. L. Deardorff, R. J. Allain, S. W. Bonham, M. H. Dancy, and J. S. Risley, "The Student-Centered Activities for Large Enrollment Undergraduate Programs (SCALE-UP) project", in *Reviews in PER: Volume 1: Research-Based Reform of University Physics*, edited by E. F. Redish and P. J. Cooney (2007).
2      C. H. Crouch and E. Mazur, "Peer Instruction: ten years of experience and results," Am. J. Phys. **69** (9), 970-977 (2001).
3      D.R. Sokoloff and R.K. Thornton, *Interactive Lecture Demonstrations, active learning in introductory physics*. (Wiley, Hoboken, NJ, 2004).
4      L.C. McDermott, P.S. Shaffer, and the University of Washington's Physics Education Group, *Tutorials in Introductory Physics*. (Prentice Hall, Upper Saddle River, NJ, 2002).
5      P.R. L. Heron, "Student performance on conceptual questions: Does instruction matter?," AIP Conf. Proc. **1513**, 174-177 (2013).
6      A. Madsen, S. McKagan, and E.C. Sayre, "Resource Letter: Research-based Assessments in Physics and Astronomy", <http://arxiv.org/abs/1605.02703>
7      L.C. McDermott and E.F. Redish, "Resource Letter: PER-1: Physics Education Research," Am. J. Phys. **67** (9), 755-767 (1999).
8      D.E. Meltzer and R.K. Thornton, "Resource Letter ALIP-1: Active-Learning Instruction in Physics," Am. J. Phys. **80** (6), 478-496 (2012).
9      R.R. Hake, "Interactive-engagement versus traditional methods: A six-thousand-student survey of mechanics test data for introductory physics courses," Am. J. Phys. **66** (1), 64-74 (1998).
10     PhysPort web site, <www.physport.org/assessments>
11     D. Hestenes, M. Wells, and G. Swackhamer, "Force Concept Inventory," Phys. Teach. **30**, 141-158 (1992).
12     R.K. Thornton and D.R. Sokoloff, "Assessing student learning of Newton's laws: The Force and Motion Conceptual Evaluation and the Evaluation of Active Learning Laboratory and Lecture Curricula," Am. J. Phys. **66** (4), 338-352 (1997).
13     S. Ramlo, "Validity and reliability of the force and motion conceptual evaluation," Am. J. Phys. **76** (9), 882-886 (2008).
14     R.K. Thornton, D. Kuhl, K. Cummings, and J. Marx, "Comparing the force and motion conceptual evaluation and the force concept inventory," Phys. Rev. ST Phys. Educ. Res. **5**, 1-8 (2009).
15     Carnegie Foundation for the Advancement of Teaching, "The Carnegie Classification of Institutions of Higher Education, 2010 edition", (Menlo Park, CA, 2011).
16     The College Board, *College Handbook 2014: All New 51st Edition*. (The College Board, New York, 2013).

17  The data are available at: [URL will be inserted by AIP].
18  In this context, the effect size means the difference between the mean scores on the pre-test and post-test divided by the pooled standard deviation. L. V. Hedges and I. Olkin, *Statistical Methods for Meta-Analysis.* (Academic Press, Orlando, FL, 1985).
19  S. Freeman, S. Eddy, M. McDonough, M.K. Smith, N. Okoroafor, H. Jordt, and M.P. Wenderoth, "Active learning increases student performance in science, engineering, and mathematics," P. Natl. Acad. Sci. U.S.A. **111** (23), 8410-8415 (2014).
20  V.P. Coletta and J.A. Phillips, "Interpreting FCI scores: Normalized gain, preinstruction scores, and scientific reasoning ability " Am. J. Phys. **73** (12), 1172-1182 (2005).
21  K. Bedard and P. Kuhn, "Where class size really matters: Class size and student ratings of instructor effectiveness," Econ. Educ. Rev. **27**, 253-265 (2008).
22  E. Mazur, *Peer Instruction: a User's Manual*. (Prentice Hall, Upper Saddle River, NJ, 1997).
23  L.C. McDermott, "Oersted Medal Lecture 2001: "Physics Education Research -- The Key to Student Learning"," Am. J. Phys. **69** (11), 1127-1137 (2001).
24  American Modeling Teachers Association web site, <http://modelinginstruction.org >
25  E. Brewe, V. Sawtelle, L.H. Kramer, G.E. O'Brien, I. Rodriguez, and P. Pamelá, "Toward equity through participation in Modeling Instruction in introductory university physics," Phys. Rev. ST Phys. Educ. Res. **6**, 010106 (2010).
26  J.M. Wilson, "The CUPLE physics studio," Phys. Teach. **32** (9), 518-523 (1994).
27  V.P. Coletta, J.A. Phillips, and J.J. Steinert, "Interpreting force concept inventory scores: Normalized gain and SAT scores," Phys. Rev. ST Phys. Educ. Res. **3**, 010106 (2007).
28  National Center for Education Statistics, Digest of Education Statistics, Table 154, <https://nces.ed.gov/programs/digest/d11/tables/dt11_154.asp>

**Figure Captions**

Fig. 1: Normalized gain for traditional lecture (TRAD) and interactive engagement (IE) for the FCI and the FMCE. Error bars are one standard error of the mean. The number at the bottom of each bar is the number of classes represented by that bar.

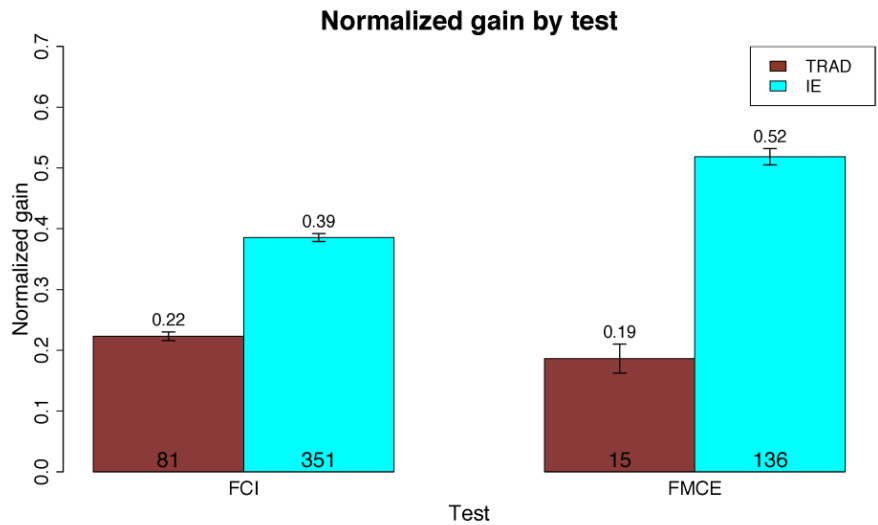

Fig. 2: Histograms of normalized gain for traditional lecture (TRAD) and interactive engagement (IE): FCI (left) and FMCE (right).

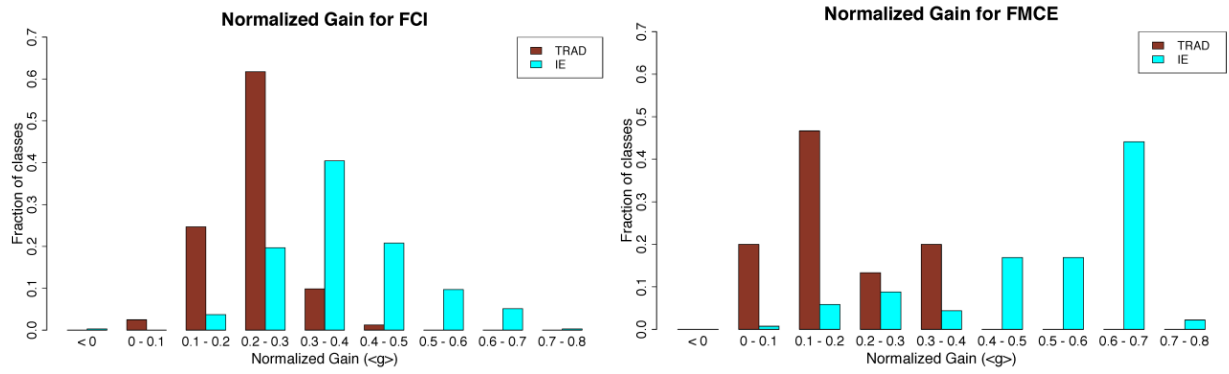

Fig. 3: Normalized gain vs class size: FCI (left) and FMCE (right). The saturation of each dot is proportional to the number of classes that have that same gain and size.

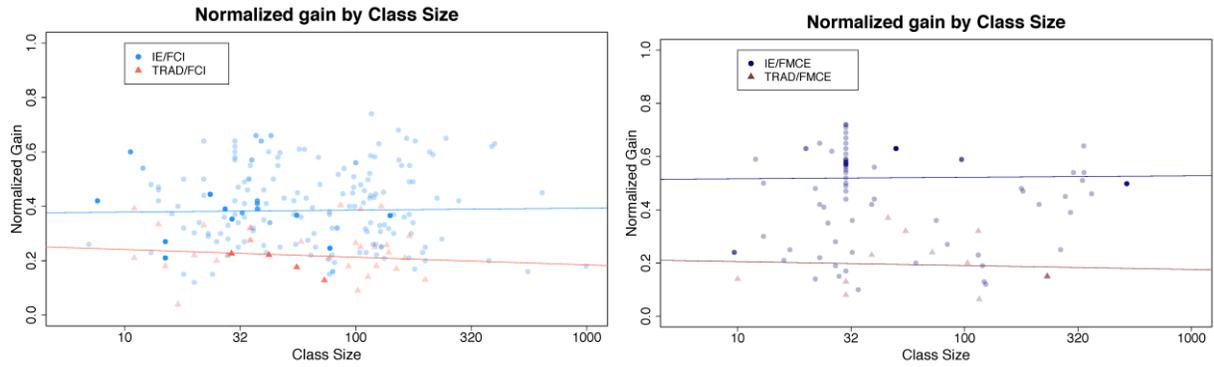

Fig. 4: Normalized gain by type of institution for the FCI. The number at the bottom of each bar is the number of classes represented by that bar. The bar that does not include error bars represents only two classes, which came from papers that did not report uncertainties.

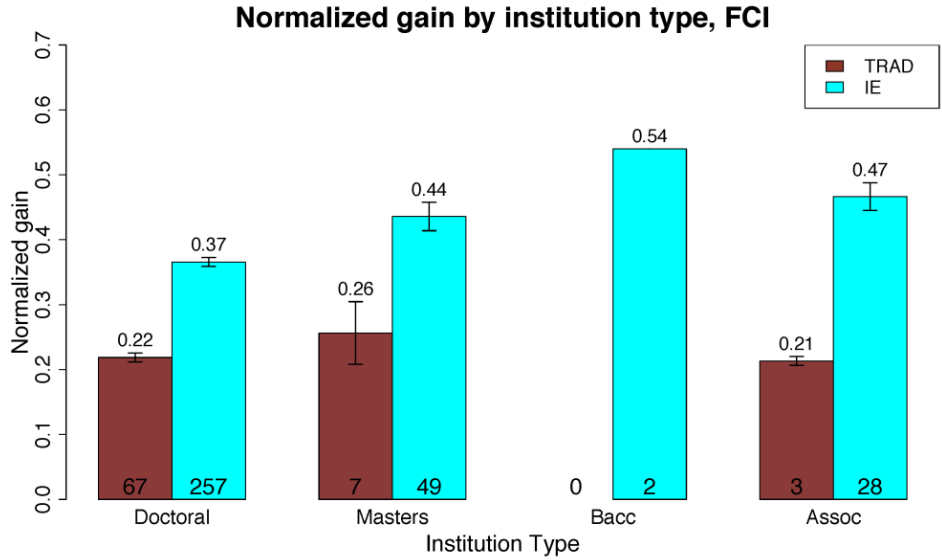